\documentclass{JHEP}
\preprint{ KUL-TF-99/24}
\title{NS Fivebranes in Type 0 String Theory\thanks{
Work supported by the European Commission TMR programme ERBFMRX-CT96-0045.}}
\author{Ben Craps\thanks{Aspirant FWO, Belgium} \ and Frederik Roose \\
Instituut voor theoretische fysica \\
Katholieke Universiteit Leuven, B-3001 Leuven, Belgium \\
E-mail:
\email{Ben.Craps,Frederik.Roose@fys.kuleuven.ac.be}}
\abstract{
The massless degrees of freedom of type 0 NS5-branes are derived. A
non-chiral, purely bosonic spectrum is found in both type 0A and 0B. This 
non-chirality is  confirmed by a one-loop computation in the bulk. Some puzzles
concerning type 0B S-duality are pointed out in this context. An 
interpretation of the spectra in terms of ``type 0 little strings'' is proposed.
}

\newcommand{\eq}[1]{Eq.~(\ref{#1})}

\def\beq{\begin{equation}}
\def\eeq{\end{equation}}
\def\beqa{\begin{eqnarray}}
\def\eeqa{\end{eqnarray}}

\newcommand{\EQ}{\begin{equation}}
\newcommand{\EN}{\end{equation}}
\newcommand{\bea}{\begin{eqnarray}}
\newcommand{\ena}{\end{eqnarray}}

\renewcommand{\thefootnote}{\fnsymbol{footnote}}
\def\one{{\hbox{ 1\kern-.8mm l}}}

\newlength{\bredde}
\def\slash#1{\settowidth{\bredde}{$#1$}\ifmmode\,\raisebox{.15ex}{/}
\hspace*{-\bredde} #1\else$\,\raisebox{.15ex}{/}\hspace*{-\bredde} #1$\fi}

\newsavebox{\uuunit}
\sbox{\uuunit}
    {\setlength{\unitlength}{0.825em}
     \begin{picture}(0.6,0.7)
        \thinlines
        \put(0,0){\line(1,0){0.5}}
        \put(0.15,0){\line(0,1){0.7}}
        \put(0.35,0){\line(0,1){0.8}}
       \multiput(0.3,0.8)(-0.04,-0.02){12}{\rule{0.5pt}{0.5pt}}
     \end {picture}}

\newsavebox{\zzzbar}
\sbox{\zzzbar}
  {\setlength{\unitlength}{0.9em}
  \begin{picture}(0.6,0.7)
  \thinlines
  \put(0,0){\line(1,0){0.6}}
  \put(0,0.75){\line(1,0){0.575}}
  \multiput(0,0)(0.0125,0.025){30}{\rule{0.3pt}{0.3pt}}
  \multiput(0.2,0)(0.0125,0.025){30}{\rule{0.3pt}{0.3pt}}
  \put(0,0.75){\line(0,-1){0.15}}
  \put(0.015,0.75){\line(0,-1){0.1}}
  \put(0.03,0.75){\line(0,-1){0.075}}
  \put(0.045,0.75){\line(0,-1){0.05}}
  \put(0.05,0.75){\line(0,-1){0.025}}
  \put(0.6,0){\line(0,1){0.15}}
  \put(0.585,0){\line(0,1){0.1}}
  \put(0.57,0){\line(0,1){0.075}}
  \put(0.555,0){\line(0,1){0.05}}
  \put(0.55,0){\line(0,1){0.025}}
  \end{picture}}
\newcommand{\Zbar}{\mathord{\!{\usebox{\zzzbar}}}}

\begin{document}

\renewcommand{\thefootnote}{\arabic{footnote}}
\setcounter{footnote}{0}
\section{Introduction}
Type 0 string theories \cite{dixharv} have recently attracted a lot of attention. They have been used to
study the non-supersymmetric field theories living on their D-branes \cite{list}. 
Moreover, Bergman and Gaberdiel
\cite{berggab} 
have conjectured that they are supersymmetry breaking orbifolds of M-theory, thus including them in the 
web of string dualities. For instance, certain circle compactifications of type 0 and type II are proposed to be
T-dual.
Other predictions of the conjectured duality are that type 0B string theory exhibits S-duality and that the
type 0 theories contain non-perturbative fermionic states. 
In Ref.~\cite{imamura} it has been studied how branes transform under type 0/type II T-duality.  

D-branes in type 0 theories have been extensively studied \cite{list}. Since their dynamics is
governed by open strings ending on them, D-branes can be studied in perturbative string theory. This
is not the case for solitonic (NS) fivebranes, which can be studied by examining the zero
modes of the
corresponding classical supergravity solutions \cite{CHS}. Each such zero mode corresponds to a massless
field on the fivebrane worldvolume.
 
In this paper we derive the massless field content of type 0 NS5-branes. Rather than doing the zero
mode
analysis directly for type 0A/B NS5-branes, we T-dualize them to Kaluza-Klein monopoles of type 0B/A, where
the zero modes can be read off from the ten-dimensional bulk fields and the cohomology of the Euclidean
Taub-NUT space. For both NS5-branes we find a non-chiral spectrum: an unrestricted antisymmetric tensor and
six scalars in 0A and two gauge fields and four scalars in 0B.
  
In type IIA the spectrum of the NS5 is chiral and anomalous. The gravitational anomaly on the NS5 must be
cancelled by anomaly inflow \cite{CH} from the bulk. The anomaly-cancelling bulk term \cite{duff} can be 
shown to be a string one-loop effect \cite{vafawitten}. The fact that we find a non-anomalous spectrum on 
type 0A NS5-branes predicts that such an anomalous bulk term is not present in type 0A. Indeed, we will
see that the type 0A GSO projection does not allow a CP-odd one-loop term.

Having identified the rather unusual NS5-brane spectrum in type 0B, one naturally wonders what the NS5
transforms into under the conjectured type 0B S-duality. Following Ref.~\cite{berggab} it seems natural to
propose that the dual object is a bound state of an electric and a magnetic D5-brane. However, we will point out
that this proposal gives rise to some puzzles.

Finally, we propose an interpretation of the NS5-brane spectra in terms of ``type 0 little strings''. This
interpretation is suggested by the doubling of the gauge degrees of freedom on the NS-branes, which is
reminiscent of (and, in fact, closely related to) the doubling of the Ramond-Ramond spectrum in going from type
II to type 0.

\section{Type 0 strings and D-branes} 

In the Neveu-Schwarz-Ramond formulation, type II string theories are obtained 
by imposing independent GSO projections on the left- and right-moving parts. 
This amounts to keeping the following (left,right) sectors: 
\beqa
{\rm IIB:}&&(NS+,NS+)\ ,~~(R+,R+)\ ,~~(R+,NS+)\ ,~~(NS+,R+)~;\nonumber\\ \label{IIspectr}
{\rm IIA:}&&(NS+,NS+)\ ,~~(R+,R-)\ ,~~(R+,NS+)\ ,~~(NS+,R-)~,
\eeqa
where for instance R$\pm$ is the Ramond sector projected with 
$P_{\rm GSO}=(1 \pm (-)^F)/2$,
$F$ being the world-sheet fermion number.

The type 0 string theories contain instead the following sectors:
\beqa
{\rm 0B:}&&(NS+,NS+)\ ,~~(NS-,NS-)\ ,~~(R+,R+)\ ,~~(R-,R-)~; \nonumber\\ \label{0spectr}
{\rm 0A:}&&(NS+,NS+)\ ,~~(NS-,NS-)\ ,~~(R+,R-)\ ,~~(R-,R+)~.
\eeqa
These theories do not contain bulk spacetime fermions, which would have to
come from ``mixed'' (R,NS) sectors.
The inclusion of the NS-NS sector with odd fermion numbers means that the 
closed string tachyon is not projected out. 
The third difference with type II theories is that the R-R spectrum is 
doubled.

As discussed in the introduction, the spectrum of type 0 theories contains two 
($p+1$)-form R-R potentials for each even (0A) or odd (0B) $p$. We will
denote these by $C_{p+1}$ and $C'_{p+1}$, where the unprimed potentials are the
ones present in type II. 
For our purposes, more convenient combinations are
\beq \label{C+-}
(C_{p+1})_\pm=\frac{1}{\sqrt{2}}(C_{p+1}\pm C'_{p+1})~.
\eeq 
For $p=3$ these are the electric ($+$) and magnetic ($-$) potentials 
\cite{9811035}. We will adopt this terminology also for other values of $p$. 

There are four types of ``elementary'' D-branes \cite{gaberdiel, 9811035} for each $p$: an electric and a 
magnetic one ({\it i.e.} charged under $(C_{p+1})_\pm$), and the
corresponding antibranes. 
In  Ref. \cite{9811035} the interaction energy of two identical parallel 
($p+1$)-branes was derived by computing the relevant
cylinder diagram in the open string channel, analogously to the Polchinski 
computation~\cite{Pol} in type II. Isolating, via modular transformation, the 
contributions due to the exchange of long-range fields in the closed string 
channel, it is found on the one hand that the tension of these branes is a 
factor $\sqrt 2$ smaller than for type II  branes. 
On the other hand, the R-R repulsive force between two like branes  
has the double strength of the graviton-dilaton attraction \cite{9811035}; 
thus the type 0 branes couple to the corresponding R-R potentials 
$(C_{p+1})_\pm$ with the same charge as the branes in type II couple to the  
potential $C_{p+1}$.

The open strings stretching between two like branes are bosons, just like the 
bulk fields of type 0. However, fermions appear from strings between an 
electric and a magnetic brane \cite{gaberdiel, BCR}.
\section{NS fivebrane spectra}
In type II string theories the massless spectrum on NS5-branes can be derived via an analysis of the 
zero modes of the
classical supergravity NS5-brane solutions \cite{CHS}. We will find it convenient to do the analysis for the
T-dual objects, since there the geometrical interpretation is manifest. 

Under T-duality the type IIA/B NS5-branes are mapped to type IIB/A Kaluza-Klein (KK) monopoles. A KK monopole
is described by a supergravity solution with six worldvolume directions and a Euclidean Taub-NUT (ETN) 
metric in 
the transverse space. The massless fields on the KK monopole correspond to the zero modes of the bulk
supergravity fields on ETN. The normalizable harmonic forms on ETN consist of one selfdual two-form. 
In the NS-NS sector there are
three zero modes from broken translation invariance and one scalar from the NS two-form. These four scalars
correspond to the translation zero modes on the corresponding NS5-branes. The R-R fields give additional
bosonic zero modes: on the IIB KK monopole the R-R two-form gives rise to a scalar and the selfdual%
\footnote{
By abuse of language we term the potential forms (anti-)selfdual, according to the \mbox{(anti-)}selfdual
 nature of their respective field strengths.} 
R-R
four-form potential leads to a selfdual two-form; the IIA KK monopole has a vector coming from the R-R
three-form. In addition there are fermionic zero modes from broken supersymmetry: two fermions with opposite
chirality for the IIA KK monopole and of the same chirality for IIB.

All in all, this leads to an ${\cal N}=(2,0)$ tensor multiplet on the IIA NS5 and an
${\cal N}= (1,1)$ vector multiplet on the IIB NS5.

Repeating this analysis for type 0 NS5-branes, it is clear that the fermionic zero modes disappear (there are
no fermions in the bulk). Nothing changes as far as the NS-NS zero modes are concerned, 
but from the R-R fields we find extra zero
modes from the doubled R-R spectrum in the bulk. On the 0B NS5 there is an extra vector (compared to the
IIB NS5), whereas the 0A NS5  gets an extra scalar and an anti-selfdual tensor. In particular the spectrum is
non-chiral on both branes.    
\section{Absence of anomalies}
At one string loop the type IIA tree level supergravity action is supplemented with the
Wess-Zumino type term $\int B \wedge X_8(R)$ coupling the NS-NS two-form $B$ to four gravitons.
$X_8$ is a quartic polynomial in the spacetime curvature.
The original derivation of this one-loop term \cite{vafawitten} provided a non-trivial consistency check
for six-dimensional heterotic-type II duality. Apart from that, this specific piece of the
action is also responsible for gravitational anomaly cancellation on IIA NS
fivebranes \cite{duff}. 

Let us first highlight some aspects of the 
direct calculation in type II, which will enable us to draw conclusions on the analogous 
type 0 amplitude almost effortlessly. First, as the interaction contains the ten-dimensional
$\epsilon$ symbol, only odd spin structures on the torus can contribute,
{\it i.e.} either the left-moving or
the right-moving fermions but not both, have to be in the odd spin structure. 
Furthermore, the even spin structures are summed over to achieve modular invariance. 
From a different perspective this may also be seen as performing the GSO projection on the
closed string states in the loop \cite{SW}. In a Hamiltonian framework the torus vacuum diagram is 
written as a
trace over the closed string Hilbert space, which decomposes in four different sectors as in
\eq{IIspectr}. Chiral traces over these sectors can be expressed in terms of traces over fermion sectors
with periodic (P) or antiperiodic (A) boundary conditions in the ($\sigma$,$\tau$) worldsheet 
directions:
\begin{eqnarray*}
{\rm Tr}_{(NS,-)} &=& \frac 1 2 \left((A,A) - (A,P) \right) \ ; \\
{\rm Tr}_{(NS,+)} &=& \frac 1 2 \left((A,A) + (A,P) \right) \ ; \\
{\rm Tr}_{(R,-)} &=& \frac 1 2 \left((P,A) - (P,P) \right) \ ; \\
{\rm Tr}_{(R,+)} &=& \frac 1 2 \left((P,A) + (P,P) \right) \ . 
\end{eqnarray*}
Notice that the odd $(P,P)$ spin structure only occurs in the Ramond sector. From the
GSO-projected spectrum in \eq{IIspectr} only (R,NS) and (R,R) sectors contain (odd,even) pieces,
yielding 
\begin{equation}\label{IIspinsum}
\frac 1 4 \left((P,P),(A,A) + (A,P) + (P,A) \right)\ .
\end{equation}
The right-moving combination is modular invariant, and in fact vanishes as a
consequence of the ``abstruse identity''. If one were to calculate the amplitude for one
B-field and four gravitons, one would insert vertex operators in the appropriate pictures.
Still, the right-moving even structures would be summed over as in \eq{IIspinsum}. In contrast with the
partition function however, in this case a non-vanishing result would be found so as not to
contradict the alternative calculation in Ref. \cite{vafawitten}. 
It is also argued there that in type IIB the
(odd,even) and (even,odd) contributions cancel out whereas they add in type IIA. So only in the
latter is the probed interaction present.

This difference between type IIA and IIB could have been inferred from their respective NS
fivebrane worldvolume spectra, a chiral one in type IIA and a non-chiral one in type IIB. The
chiral spectrum suffers from a gravitational anomaly in six dimensions, which is neatly cancelled
by the standard anomaly inflow \cite{CH} from the bulk through the above derived coupling \cite{duff}. 

Let us repeat the argument for the one-loop correlator in type 0 string
theories. Now it is only (R,R) sectors that contribute to the (odd,even) piece in the one-loop
partition function, yielding
\begin{equation}\label{0spinsum}
\frac 1 4 \left( (P,P)-(P,P), (P,A) \right) 
\end{equation}
both for 0A and 0B. As in type II one may now wish to insert vertex operators. However, it is
obvious that the obligatory sum over spin structures as in \eq{0spinsum} 
now yields a vanishing result for
the (odd,even) part by itself. Analogously, the (even,odd) part will vanish by itself. 
Whence the absence of the anomaly-cancelling term in both type 0A and 0B.

Reversing the anomaly inflow argument in type 0A the  selfdual  two-form must be
supplemented with an anti-selfdual two-form, in order not to give rise to gravitational
anomalies. This nicely agrees with the above derived unconstrained two-form on the NS
fivebrane. From T-duality an additional vector (compared to the IIB NS5-brane)
has to be expected on the type 0B NS5-brane.

\section{S-duality}
\paragraph{Strings} In type IIB string theory the fundamental string is mapped to the D-string under S-duality.
The Green-Schwarz light-cone formulation  of the fundamental type IIB string involves eight spacetime
bosonic fields and sixteen spacetime fermionic ones. All fermions transform in the ${\bf 8_S}$ representation
of the transverse SO(8) rotation group. Upon quantization, their zero modes generate the 256-fold degenerate
groundstate, transforming in the $({\bf 8_V}+{\bf 8_C})\otimes({\bf 8_V}+{\bf 8_C})$ of SO(8). 

The excitations of a D1-brane can be described in string perturbation theory by quantizing the open strings
beginning and ending on it. Doing so, the NS sector gives eight massless bosons and the R sector sixteen
massless fermions.

Note that these massless excitations of both the fundamental string and the D-string can be interpreted as
Goldstone modes for broken translation invariance and broken supersymmetry.

It is believed that the IIB D-string at strong (fundamental) string coupling is the same object as a
fundamental IIB string at weak coupling. The evidence for this conjecture is largely based on supersymmetry.
For one thing, the ground state of the D-string is BPS, so that it can safely be followed to strong coupling
and compared to the ground state of the weakly coupled fundamental string. 

Type 0B string theory is not supersymmetric, which complicates the analysis of its strong coupling limit
enormously. Let us nevertheless try to proceed.

The fundamental type 0B string can be obtained from the fundamental type IIB string by performing a
$(-1)^{F_s}$ orbifold, where $F_s$ is the spacetime fermion number. In the untwisted sector of this
orbifold the massless states in the $({\bf 8_V}\otimes{\bf 8_V})+({\bf 8_C}\otimes{\bf 8_C})$ survive. The
light modes in the twisted sector are a singlet tachyon and the extra massless R-R states in the 
${\bf 8_S}\otimes{\bf 8_S}$.

Bergman and Gaberdiel have 
conjectured \cite{berggab} that the  fundamental type 0B string should be S-dual to a
bound state an electric and a magnetic D1-brane. This proposal immediately raises two questions. First, when
one quantizes the open strings on a coincident electric-magnetic D-string pair, one finds twice as many
massless excitations as on a fundamental string: sixteen bosons and thirty-two fermions. This would
lead to way too many states of the D-string pair compared to the fundamental string. How could the extra
degrees of freedom disappear? Second, assuming that we have found a way to get rid of the superfluous degrees
of freedom, how does the $(-1)^{F_s}$ projection appear on the D-string pair?

Concerning the first question, let us just note that this
naive mismatch is no immediate contradiction: to compare the D-string pair to the weakly coupled fundamental 
string, one should take the former to strong (fundamental string) coupling. Along the way, the excitation 
spectrum may well change. As to the second question, the $(-1)^{F_s}$ projection might be related to a gauging 
of a $\Zbar_2$ subgroup of the gauge symmetry on the D-string pair. However, it looks hard to make these ideas
more precise.

\paragraph{Fivebranes} Type IIB NS5-branes are S-dual to D5-branes. These objects have the same light
excitations (the reduction of an ${\cal N}=1$, D=10, U(1) vector multiplet to D=6). As in the case of IIB
strings, part of the spectrum is protected by supersymmetry.

Given the story for strings, it looks reasonable to propose that the type 0B NS5-brane is S-dual to
an electric-magnetic D5-brane pair. This raises two puzzles. 

First, how do the spectra match? Is the doubled gauge spectrum of type 0B NS5-branes at weak coupling related 
to the two
gauge fields on the D5-brane pair? This is not clear, since to compare the two objects one has to take one of
them to strong coupling, where it is unclear what happens. For one thing, the doubling of the gauge field on the
type 0B NS5 (compared to the IIB NS5) is related to the doubled R-R spectrum of the bulk of type 0A, as shown in
Section~3. The latter doubling is not expected to persist at strong coupling \cite{berggab}.
Are there fermionic zero modes on NS5-branes? We have not found any from a zero mode analysis of the bulk
gravity theory, but this does not exclude that they could appear in a non-perturbative way. This would be
analogous to the (not completely understood) case of the fundamental type 0 string, where it looks impossible to
interpret the fermion zero modes as zero modes of bulk fields.

Second, do type 0B NS5-branes carry anomalous gravitational couplings, which are known to be carried by type 0B
D-branes \cite{BCR}? Since these couplings are related to anomalies and thus to topology, we hope that their presence or
absence could be invariant under continuously changing the coupling from zero to infinity. If this is the case,
answering this question could provide a test of type 0B S-duality. The rest of this section will be devoted to
analysing this question.

We first remind the reader of the familiar story in type IIB. Consider, for instance, the intersection of two
D5-branes on a string. The open strings from one brane to the other give rise to chiral fermions living on the
intersection. These chiral fermions lead to gauge and gravitational anomalies, which are cancelled by anomaly
inflow from the branes into the intersection \cite{GHM}. For this to happen, the D5-branes need, amongst others, 
a
\beq
C_2\wedge p_1
\eeq     
term in their Wess-Zumino action, where $C_2$ is the R-R two-form potential and $p_1$ the first Pontrjagin class
of the tangent bundle of the D5-brane worldvolume. The presence of this term has been checked by an explicit
string computation \cite{benfred}. Now, type IIB S-duality implies the presence of chiral fermions on the
intersection of two NS5-branes on a string, and thus a
\beq
B\wedge p_1
\eeq
term on the worldvolumes of each NS5-brane ($B$ is the NS-NS two-form). Indeed, the following argument%
\footnote{In fact, this is how the $C_2\wedge p_1$ coupling on D5-branes was first discovered.}
for the
presence of this term was given in Ref.~\cite{bersadvaf}. Under T-duality in a transverse direction the 
NS5-brane is mapped to a Kaluza-Klein monopole in type IIA. This object is described by a Euclidean Taub-NUT
metric in the transverse space.  Since $p_1({\rm ETN})\neq 0$ and the $B\wedge X_8$ term discussed 
in
the previous section contains a piece proportional to $B\wedge (p_1)^2$, reducing this term 
gives rise to the required $B\wedge p_1$ term on the IIA Kaluza-Klein
monopole and thus on the IIB NS5-brane.

What about type 0B? It has been argued \cite{BCR} that chiral fermions and thus anomalies are present on the
intersection of electric and magnetic D5-branes. Hence, type 0 D5-branes have the $p_1$ couplings in their
Wess-Zumino actions (this has also been checked via a string computation \cite{BCR}). We expect that the number
of chiral fermion zero modes on these intersections cannot change under a continuous increase of the string 
coupling. On the other hand, in this paper we have argued that there is no $B\wedge X_8$ term in type 0A, so we
expect no  $B\wedge p_1$ coupling of type 0B NS5-branes, and thus no chiral fermions on intersections of
NS5-branes. If this argument is correct, it could be a problem for type 0B S-duality.
       
\section{A little string interpretation}
In type II string theories the fivebrane worldvolume theories have been conjectured to be
properly described by ``little strings''\cite{little}. 
These non-critical closed string theories have 
hitherto remained mysterious, although some qualitative features may be understood. 
We briefly recall some of these, focusing on the type IIB case. However, we
prefer first to treat the more familiar example of a fundamental string, providing a guiding
reference when dealing with the little strings. 

A fundamental string solution in supergravity breaks half of the 32 bulk supersymmetries. These
broken symmetries give rise to fermion zero modes on the string, of which 8 are left-moving 
and 8 right-moving. They are identified with the Green-Schwarz space-time fermion fields on
the worldvolume. Upon quantization of these fermions 128 bosons and 128 fermions are found,
together building the on-shell ${\cal N}$ = 2 supergravity multiplet. This multiplet is also
found from the RNS superstring, so the picture is consistent. In the latter formulation it is
not hard to see that the NS-NS sector fields together with, say, the R-NS sector build an
${\cal N}$ = 1 supergravity multiplet. Likewise, the NS-R and R-R sector are combined 
into one multiplet of ${\cal N}$ = 1. 

Let us move on now to little strings.
One way to derive properties of little strings in type IIB is to consider a gauge instanton on
a D5 brane. 
Half of the bulk supersymmetries are unbroken by the D5-brane. Sixteen real supercharges are thus found, 
obeying a ${\cal N}$=(1,1) 
superalgebra. Of these, half are broken by the instanton. On the macroscopic string there are
thus eight fermionic modes, four left-moving and four right-moving ones. Upon quantization of these
zero modes one finds 8 bosonic and 8 fermionic states. They correspond to the on-shell degrees
of freedom of one ${\cal N}$=(1,1) vectormultiplet.%
\footnote{Since the little strings of type IIB are non-chiral, we will name them type iia little strings.} 
Upon decomposition with respect to ${\cal N}$ = 1 six-dimensional supersymmetry one
hypermultiplet and one vectormultiplet result. It is conceivable that a similar analysis 
in type IIA, if possible at all, would give a ${\cal N}$=(2,0) tensormultiplet, decomposing into
one hypermultiplet and one tensormultiplet of ${\cal N}$=1 supersymmetry. We remark that the
tensormultiplet has a bosonic content that consists of one scalar and one two-form potential
with self-dual field strength. All in all, one may wish to draw the following analogies between
little strings and type II strings. In both cases there is one universal sector, the
hypermultiplet resp. the (NS-NS + R-NS) sector. The other sector depends on whether or not the
supersymmetry in the
``bulk'' is chiral, where ``bulk'' means the fivebrane worldvolume for the little string 
and ten-dimensional spacetime for
the type II string. For the non-chiral type IIA and little type iia the remaining
bosonic fields are odd $q$-form potentials, coupling to even $p$ D$p$- resp. d$p$-branes. As to chiral
type IIB/iib the R-R sector has only even $q$-form potentials, of which one has a self-dual field
strength. In these theories only odd $p$ D$p$ or d$p$-branes occur. So this analogy, however formal,
is at least remarkable. 

As to the various fivebranes, the type IIA NS fivebrane theory is conjectured to be 
described by chiral iib little strings, whereas iia little strings would do the job for type
IIB fivebranes.

If one is willing to take the above for granted, 
one may then proceed and construct ``little type 0a/b strings'' by
mimicking the procedure for the bulk strings: spacetime fermions are removed and there is a 
doubled R-R spectrum compared to the superstring.
Concretely this would imply that the massless spectrum of little type 0a strings consists of
the bosonic contents of one hypermultiplet and two vectormultiplets, in ${\cal N}$=1 language. 
This precisely matches with the spectrum on the type 0B NS fivebrane. Little type 0b strings
would then generate the bosonic content of one hypermultiplet, one selfdual tensormultiplet
and one anti-selfdual tensormultiplet. Also here this seems to be consistent with the spectrum
of the type 0A NS fivebrane. 

\section{Conclusions}
From a zero mode analysis of type 0 Kaluza-Klein monopoles, we have derived the
massless degrees of freedom of NS5-branes in type 0A and 0B. The massless
spectrum is purely bosonic and non-chiral in both cases. The 0A NS5-brane
contains an unrestricted tensor (rather than a selfdual one) whereas two gauge fields live on
the 0B NS5-brane. 

Since the spectrum of a 0A NS5-brane is non-chiral, the anomaly-cancelling one
loop bulk term present in type IIA should be absent in type 0A. This is indeed
the case.

We have pointed out some puzzles concerning type 0B S-duality. In particular, the question is raised
whether the absence of a  $B\wedge X_8$
term in the bulk of type 0A makes the duality of NS5-branes with D5-brane
pairs impossible.  

It is tempting to interpret the spectrum we find in terms of ``type 0 little
strings''. Their relation to the usual type II ``little strings'' is very
similar of the relation between the bulk type 0 and type II strings: fermions
are projected out and the ``Ramond-Ramond'' spectrum is doubled.

\acknowledgments{
We would like to thank M.~Bill\'o, M.~Costa, M.~Gaberdiel, R.~Gopakumar, 
I.~Klebanov, A.~Sen, J.~Troost and W.~Troost for 
interesting  discussions, comments and suggestions.}


\end{document}